\documentclass[twocolumn,showpacs,preprintnumbers,amsmath,amssymb]{revtex4}

\usepackage{graphicx}
\usepackage{dcolumn}
\usepackage{bm}

\begin{document}

\preprint{}

\title{Cosmic-ray spectra of primary protons and  
high altitude muons deconvolved 
from observed atmospheric gamma rays}

\author{K. Yoshida}
\affiliation{Department of Electronic Information Systems, 
Shibaura Institute of Technology, Saitama, Japan}
\email{yoshida@shibaura-it.ac.jp}

\author{R. Ohmori}
\affiliation{Utsunomiya University, Utsunomiya, Japan}

\author{T. Kobayashi}
\affiliation{Department of Physics and Mathematics, Aoyama Gakuin University,
Sagamihara, Japan}

\author{Y. Komori}
\affiliation{Kanagawa University of Human Services, Yokosuka, Japan}

\author{Y. Sato}
\affiliation{Utsunomiya University, Utsunomiya, Japan}

\author{J. Nishimura}
\affiliation{Institute of Space and Astronautical Science,
Sagamihara, Japan}

\date{\today}

\begin{abstract}
We have observed atmospheric gamma rays from 30~GeV to 8~TeV, 
using emulsion chambers at balloon altitudes, 
accumulating the largest total exposure in this energy range to date, 
$S{\Omega}T\sim$6.66~m$^2$~sr~day. 
At very high altitudes, with residual overburden 
only a few g~cm$^{-2}$, 
atmospheric gamma rays are mainly produced by a single interaction of primary cosmic rays 
with overlying atmospheric nuclei. 
Thus, we can use these gamma rays to study the spectrum of 
primary cosmic rays and their products in the atmosphere. 
From the observed atmospheric gamma ray spectrum, 
we deconvolved the primary cosmic-ray proton spectrum, 
assuming appropriate hadronic interaction models. 
Our deconvolved proton spectrum covers the energy range 
from 200~GeV to 50~TeV, 
which fills a gap in the currently available primary cosmic-ray proton spectra. 
We also estimated the atmospheric muon spectrum above 30~GeV at high altitude 
from our gamma-ray spectrum, almost without reference to the primary cosmic rays, 
and compared the estimated flux with direct muon observations below 10~GeV. 
\end{abstract}

\maketitle

\section{\label{sec:intro}Introduction}

The primary cosmic-ray proton spectrum is one of the most important 
quantities needed to interpret cosmic-ray phenomena inside the atmosphere. 
Many observations of primary protons have been performed 
since the discovery of cosmic rays. 
Recently, proton flux measurements have attracted renewed attention, since they are needed to 
precisely estimate the absolute flux of atmospheric neutrinos, providing critical input for 
analyses of neutrino oscillations being 
performed with Super-Kamiokande~\cite{fukuda98} and similar detectors. 
In particular,
atmospheric neutrinos with higher energies, $\sim100$~GeV,  
which are detected as upward through-going muons,  
are mainly produced from the interaction of $\sim1$~TeV primary protons 
in the atmosphere. 
Similarly important are observations of the absolute flux 
of muons, which are the partners of neutrinos,
and thus provide a check on atmospheric neutrino flux estimates. 
Since atmospheric gamma rays at high altitude, with residual overburden
of a few g~cm$^{-2}$, 
are mainly produced by a single interaction of primary cosmic rays 
with overlying atmospheric nuclei, 
we can use these gamma rays to study the spectrum of 
primary cosmic rays and their products in the atmosphere. 

We have been observing primary cosmic ray electrons 
at balloon altitudes using emulsion chambers for many years, and successfully obtained the energy spectrum of electrons 
in the energy range from 
30GeV to 3TeV \cite{nishimura80,kobayashi99}. 
In the course of these electron observations, 
we have simultaneously observed atmospheric gamma rays. 
In particular, 
since gamma rays produce electromagnetic showers 
just like electrons, 
emulsion chambers are not only suitable 
to observe the gamma rays in the atmosphere, 
but measurement of the gamma ray showers is a necessary part
of the electron analysis, required
to check and calibrate the performance of the emulsion chambers. 

Atmospheric gamma rays have been observed by the BETS group 
in the energy range from a few GeV to several 10~GeV 
at mountain altitude (2.77~km) and at balloon altitudes 
($15-32$~km)~\cite{kasahara02}. 
They found that hadronic interaction simulation codes such as 
the Lund group's Fritiof V7.02, or Dpmjet3 
give results fairly consistent with the observed energy spectra and 
altitude variation of the gamma-ray flux, 
within the accuracy of their measurements. 
The JACEE group also observed high-energy gamma rays in the $3-30$~TeV range at overburden $\sim$5.5~g~cm$^{-2}$ for $120-150$~hr~\cite{jacee95}.

Atmospheric neutrinos, muons, and gamma rays are decay products of pions produced in 
the interaction of primary cosmic rays with atmospheric nuclei: 

\begin{eqnarray}
p + N & \rightarrow & 
\pi^{\pm} \rightarrow \mu^{\pm} + \nu_{\mu}(\bar{\nu}_{\mu}) 
\label{eq:decay} \\
      &             & 
\hspace*{12mm} \searrow e^{\pm} + \nu_{e}(\bar{\nu}_{e}) +
 \bar{\nu}_{\mu}(\nu_{\mu}) \hspace*{1cm} \nonumber \\
      & \rightarrow & \pi^{0} \rightarrow 2\gamma \nonumber
\end{eqnarray}
In addition to pions,  
$K$ and $\eta$ mesons are produced in the primary interactions, and also
decay into muons and gamma rays, making minor contributions to the net fluxes.

Atmospheric gamma rays are mainly the decay products of $\pi^0$. 
When we observe atmospheric gamma rays at very high altitudes, corresponding to 
residual overburdens of several g~cm$^{-2}$, 
the thickness of atmosphere above the detector 
is $\sim$0.05 nuclear mean free path or $\sim$0.1 radiation lengths (r.l.).
Thus, hadronic cascades and electromagnetic showers cannot develop 
significantly in the atmosphere
before reaching the detector. 
The observed gamma rays are almost all produced by the first interaction of
primary cosmic rays (protons, helium and heavier nuclei) 
with atmospheric nuclei. 
Therefore, using the appropriate hadronic interaction models, 
we can reliably deconvolve the primary proton flux 
from the observed spectrum of atmospheric gamma rays 
produced by $\pi^0$. This can be done in a semi-analytic way, 
because of the approximate scaling nature of 
the pion production cross section. 
In this deconvolution, we include the effects of contributions 
from primary helium and heavier nuclei, 
and also the minor contribution of gamma rays 
from decays of $\eta$ and $K$ mesons.

The proton spectra have been measured with two kinds of detectors: 
magnetic spectrometers and calorimeters. 
Although magnetic spectrometers have excellent energy resolution, 
their maximum observable energy is limited by 
the maximum detectable rigidity (MDR) to $\sim$1~TeV. 
The BESS, AMS-01 and CAPRICE instruments are typical detectors 
using magnetic spectrometers, 
and have precisely measured primary
proton spectra up to $\sim100$~GeV \cite{sanuki00,aguilar02,boezio03}. 
BESS and AMS-01 give results consistent with each other 
within an error of a few \%, while CAPRICE reports a 10~\% lower proton flux. 
Recently, 
the BESS-TeV spectrometer, with improved MDR~\cite{haino04}, 
observed primary proton spectra up to 540~GeV. 
As for calorimeters, 
electronic detectors using scintillators, as well as passive 
emulsion chambers have been used 
to observe the proton spectrum in the higher energy region. 
The first measurements of the proton spectrum in the TeV region were made 
by the PROTON satellite-borne calorimeters \cite{grigorov68}. 
These observations were followed by an ionization spectrometer 
flown at balloon altitudes 
whose energy range extend from 50~GeV to 2~TeV \cite{ryan72}. 
The proton spectrum in the energy range from 10~TeV to 1000~TeV 
has been measured by the JACEE Collaboration and  
RUNJOB Collaborations using emulsion chamber detectors
\cite[and references therein]{jacee98,runjob01,derbina05}. 
Recently, 
ATIC, which is a calorimeter with BGO scintillators, 
observed primary proton spectrum in 30~GeV-50~TeV \cite{wefel05}. 
However, more
accurate measurements of the proton spectrum are still required 
in the energy range from 100~GeV to 10~TeV. 

In detectors that directly observe primary cosmic rays, such as JACEE and RUNJOB, 
the energies of primary protons are estimated from 
electromagnetic showers produced by nuclear interactions 
within the emulsion chambers. 
Since the electromagnetic shower developed in the chamber 
is actually initiated by multiple secondary gamma rays, and 
its structure is affected by their emission angles, 
energy estimation for the primary particle is complicated. 
On the other hand, 
in the present work, we derive the energy of primary protons from observed
electromagnetic showers initiated by one single gamma ray, 
so that the estimation of proton fluxes from atmospheric gamma rays is 
simpler and therefore more reliable.

As a first approximation, 
the number of charged pions produced in hadronic interactions 
is almost two times of neutral pions. 
Hence we can estimate the production rate of charged pions from 
atmospheric gamma rays. 
Since the muons are mostly produced from the decay of charged
pions, 
we can also directly estimate the muon flux without reference to the primary
cosmic ray flux, 
correcting for the minor contribution of $\eta$ and $K$ mesons. 
More precisely, widely accepted
hadronic interaction models are used to derive the muon spectrum 
from the gamma-ray spectrum.

The MASS, CAPRICE, HEAT and BESS group have performed direct
atmospheric muon observations at various balloon altitudes 
\cite{codino97,francke99,beatty04,sanuki01,abe03}. 
The muon spectrum at various atmospheric depths 
gives important information to check the reliability of 
calculations performed to estimate the neutrino flux \cite{sanuki06}. 
However, the muon flux at large depths depends on the detailed structure of atmospheric density as a function of altitude, 
which is subject to seasonal variation and meteorological conditions. 
This brings complications and ambiguities into the muon flux estimation. 
On the other hand, 
the flux of atmospheric gamma rays at stratospheric altitudes depends only on the residual atmospheric overburden, 
which can be estimated much less ambiguously than the muon flux. 
Hence, 
an accurate measurement of the gamma-ray flux can be also used 
to calibrate the neutrino flux calculations, 
because of the relation between gamma-ray and neutrino 
production in the atmosphere as shown in the process (\ref{eq:decay}).

In this paper, 
we present measurements of the atmospheric gamma-ray spectrum 
due to hadronic interactions 
in the energy range of 30~GeV to 8~TeV, observed 
at balloon altitudes with emulsion chambers. 
We also present results from the deconvolution of primary cosmic-ray proton 
and high altitude muon spectra from our gamma-ray observations.

\section{\label{sec:atm_obs}Atmospheric gamma-ray observation}

\subsection{Balloon observations}

We have observed primary electrons with balloon-borne emulsion chambers 
in many flights between 1968 and 2001 \cite{nishimura80,kobayashi99,sato03}. 
Simultaneously, we have also observed atmospheric gamma-rays 
to check the performance of the emulsion chambers in each balloon experiment. 
The pressure altitude records for each flight correspond to residual atmospheric 
overburdens in the range from 4.0~g~cm$^{-2}$ to
9.4~g~cm$^{-2}$. 
The total cumulative effective exposure 
$S{\Omega}T$ for gamma rays is $6.66$m$^2$-sr-day, 
which is larger than any other atmospheric gamma-ray observations 
performed at balloon altitudes in the energy range of 30~GeV to 8~TeV.  
In Table \ref{tab:baln_list}, we summarize the series of experiments 
since 1968. 
Nishimura {\it et al.} (1980) have reported the spectra of primary
electrons and atmospheric gamma rays 
for the 1968---1976 observations in our previous work~\cite{nishimura80}.

\begin{table*}
\caption{\label{tab:baln_list} List of balloon flights}
\begin{center}
\begin{tabular}{lrrccr}
\hline\hline
Flight & Area & Time & Average Altitude & $S{\Omega}_{\gamma}T$ $^*$ & 
Launch Site \\
       & (m$^2$) & (min) & (g~cm$^{-2}$) & (m$^2$ sr s) &  \\
\hline
1968   & 0.05 &  380 & 6.1 & $2.381\times10^3$ & Harunomachi, Japan \\
1969   & 0.05 &  267 & 7.1 & $1.799\times10^3$ & Harunomachi, Japan \\
1970   & 0.05 & 1136 & 6.1 & $7.900\times10^3$ & Sanriku, Japan \\
1973   & 0.20 &  833 & 8.2 & $2.694\times10^4$ & Sanriku, Japan \\
1976   & 0.40 & 1526 & 4.0 & $9.425\times10^4$ & Palestine, USA \\
1977   & 0.78 & 1760 & 4.5 & $6.549\times10^4$ & Palestine, USA \\
1979   & 0.80 & 1680 & 4.9 & $1.411\times10^5$ & Palestine, USA \\
1980   & 0.80 & 2029 & 7.8 & $1.099\times10^5$ & Palestine, USA \\
1984   & 0.20 & 576  & 9.2 & $7.106\times10^3$ & Sanriku, Japan \\
1985   & 0.40 & 940  & 9.4 & $1.324\times10^4$ & Sanriku, Japan \\
1988   & 0.20 & 647  & 7.1 & $3.929\times10^3$ & Uchinoura, Japan \\
1996   & 0.20 & 2092 & 4.6 & $6.497\times10^4$ & Sanriku, Japan \\
1998   & 0.20 & 1178 & 5.6 & $3.638\times10^4$ & Sanriku, Japan \\
2001   & 0.20 & 1108 & 5.5 & $1.096\times10^2$ & Sanriku, Japan \\
\hline
\multicolumn{6}{l}{$^*$ Effective $S{\Omega}_{\gamma}T$ for the gamma-ray
 observations. }
\end{tabular}
\end{center}
\end{table*}

\subsection{Detector}

\begin{figure}[h]
 \begin{center}
  \includegraphics[width=80mm]{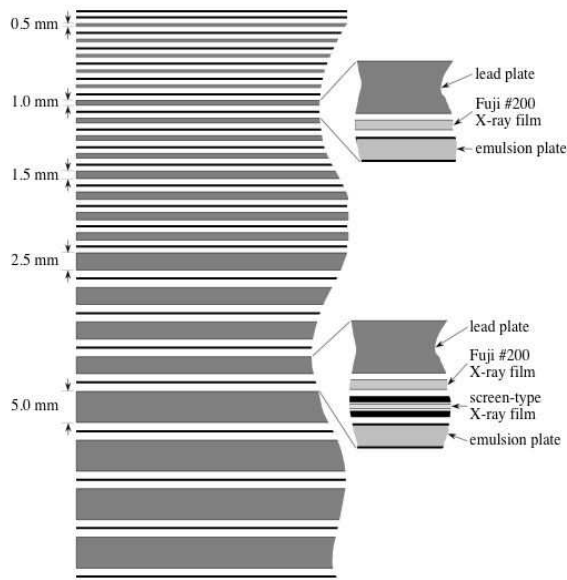}
 \end{center}
 \caption{\label{fig:ECCdesign}
Typical configuration of the emulsion chamber in cross-sectional drawing 
from side view. }
\end{figure}

Emulsion chambers consist of nuclear emulsion plates 
and lead plates, which are stacked alternately. 
Nuclear emulsion plates sample 
the development of electromagnetic showers produced in the lead plates. 
X-ray films are also inserted to allow rapid, naked-eye scanning for high
energy showers, 
which produce dark spots in the film. 
The threshold energy for shower detection depends on the background and 
the sensitivity of the films; 
some details of the detection threshold for various films 
are described later. 
Figure \ref{fig:ECCdesign} shows a typical 
emulsion chamber configuration. 
Detailed configurations and performance are described 
in Nishimura {\it et al.} (1980) \cite{nishimura80} and 
Kobayashi {\it et al.} (1999) \cite{kobayashi99}.

In emulsion chambers, 
it is possible to measure the location of shower tracks 
in each emulsion plate with a precision of $\sim$1~$\mu$m. 
Because of this high position resolution, 
we can inspect the shower starting points in detail and 
unambiguously distinguish showers due to electrons, gamma rays, 
and other hadronic interaction events \cite{nishimura80}. 
By inspecting various specific features of those events, 
the rejection power for protons misidentified as electron
candidates is found to be 
as large as 10$^5$ \cite{kobayashi99}. 

We measure the shower particles within a circle of 100~$\mu$m 
radius from shower axis, 
which means that we select the shower particles with higher energy,  
which have suffered less multiple scattering in the chamber. 
Hence, the number of the shower particles drops off faster than 
for all shower particles,  
and the shower maximum appears in $\sim$6~r.l. for 1~TeV electrons, 
while the maximum of the total number of shower particles appears 
in $\sim$12 r.l. for 1~TeV electrons. 
This means that we can determine the energy of higher energy 
incident electrons with thinner detector. 
The typical size and thickness of the detector 
are 40~cm $\times$ 50~cm, and 8~cm ($\sim$9~r.l.), respectively. 
Thus the emulsion chamber has the advantage of a large effective area combined 
with a wide field of view compared with other detectors. 
Because of the simple configuration of the detector, 
we can estimate the geometrical factor ($S{\Omega}$) 
very accurately, a task
which is difficult for some electronic detectors.

\begin{figure}[h]
 \begin{center}
  \includegraphics[width=80mm]{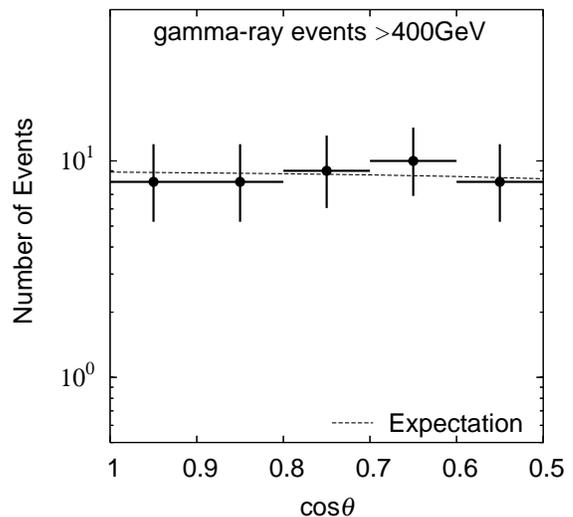}
 \end{center}
 \caption{\label{fig:g400GeVang}
Zenith angle distribution for atmospheric gamma rays beyond 400~GeV.}
\end{figure}

For electron observations,
the effective geometrical factor is given by 
\begin{equation}
S{\Omega}_e = 2{\pi}S{\eta}\int_0^{\theta_0} {\cos}{\theta}{\sin}{\theta}
d{\theta} = {\pi}S{\eta}{\sin}^2{\theta_0}, 
\end{equation}
where ${\theta_0}$ is the upper limit of incident angles and 
$\eta$ is the efficiency of events that pass through the top and bottom 
emulsion plates, the so-called ``edge effect''. 
On the other hand, 
for atmospheric gamma-ray observations, 
the effective geometrical factor is given by 
\begin{equation}
S{\Omega}_{\gamma} = 2{\pi}S{\eta}\int_0^{\theta_0} {\sin}{\theta}
d{\theta} = 2{\pi}S{\eta}(1-{\cos}{\theta_0}). 
\end{equation}
Here, we corrected $S{\Omega}_{\gamma}$ to obtain the vertical flux, 
because the gamma rays have a $1/{\cos}{\theta}$ enhancement 
relative to isotropic primary cosmic rays,  
proportional to the path length in the overlying atmosphere 
at balloon altitudes. 
Figure~\ref{fig:g400GeVang} shows the zenith angle distribution observed with 
emulsion chambers, compared with expectation calculated taking into account 
small corrections due to the elongation of the path length 
in the overlying air for large zenith angles, 
which affect the amount of absorption of the electrons and gamma rays. 
In the typical case of $\theta_0 = 60^{\circ}$ and $S=0.40{\times}0.50$~m$^2$, 
$S{\Omega}_{\gamma}$ is 0.52~m$^2$~sr with $\eta$ of 0.82 
for the chamber thickness of 8.0~cm. 
For the comparison of emulsion chambers with other detectors, 
we summarize the effective geometrical factor $S{\Omega}_{e}$ 
with efficiency 
for primary electron experiments in Table~\ref{tab:somega_e}. 
We can see how the emulsion chamber is efficient for observation of
low-flux components of cosmic rays such as primary electrons 
and atmospheric gamma rays.

\begin{table}[h]
\caption{\label{tab:somega_e}
Examples of effective geometrical factors with efficiency 
of the electron detectors.} 
\begin{center}
\begin{tabular}{lcr}
\hline\hline
 Detector  &\ \  $S{\Omega}_e$ (m$^2$sr)\ \  & reference \\
\hline
 T.R.D. & 0.11 & \cite{tang84} \\
 MASS-91 & 0.018 & \cite{grimani02} \\
 CAPRICE94 & 0.016 & \cite{boezio00} \\
 HEAT & 0.012 & \cite{duvernois01} \\
 BETS & 0.032 & \cite{torii01} \\
 AMS-01 & 0.10 & \cite{aguilar02} \\
 ECC & 0.38$^{*}$ & \cite{nishimura80} \\
\hline
\multicolumn{3}{l}{ $^*$ $S{\Omega}_e$ in the typical emulsion chamber.} \\
\end{tabular}
\end{center}
\end{table}

\subsubsection{Scanning method}

Since high energy electro-magnetic showers above several 100~GeV 
leave dark spots on X-ray films, 
we can detect these showers with the naked eye by scanning the X-ray films. 
We locate the corresponding tracks in the adjacent emulsion plate 
using microscopes, 
trace them back through the stack to the cascade starting point, 
and identify the incoming particle. 
The validity of particle identification 
in our experiments is also checked by comparison with the expected 
zenith angle distributions and 
shower starting point distributions, 
as described in our previous work \cite{nishimura80, kobayashi99}. 
We picked up events with zenith angle less than
60$^{\circ}$. 
The detection threshold of the X-ray film, 
although it depends on the accumulated background tracks and fog in the film, 
is $\sim$500~GeV for Sakura type-N X-ray film used before 1984, 
$\sim$800~GeV for Fuji \#200 X-ray film and 
$\sim$200~GeV for screen type X-ray films such as 
Fuji G8-RXO, G12-RXO, HR8-HA30, HR12-HA30 used on and after 1984 
\cite{kobayashi91}.

To detect the electro-magnetic showers below a few hundred GeV, 
the emulsion plates had to be directly scanned using microscopes. 
Although the emulsion chamber is the only detector which has succeeded 
in observing 
cosmic-ray electrons and gamma rays in the TeV region, 
it was difficult to observe electrons or gamma-rays 
below a few hundred GeV 
because of the tedious work required in manual microscope scanning 
of the emulsion plates. 
In order to overcome this difficulty tracing low energy showers, 
in 2001 we started to apply the automatic scanning method 
which has been successfully developed by the Nagoya group 
to observe $\nu_{\tau}$ events with nuclear emulsions \cite{eskut97}. 
A fully automatic emulsion scanning system developed by the Nagoya group, 
called ``Track Selector",
consists of a motor-driven microscope XYZ stage with a controller, 
a fast read-out CCD and special hardware for 3D image processing,. 
It takes the images of 16 ``slices", or layers within an emulsion plate, 
each of 512$\times$512 pixels, corresponding to a field of view of 
147$\times$106~$\mu$m$^2$, 
and then outputs position and angle data for each track found in 
the emulsion plate. 
It takes approximately 16~hr to perform general scanning over an area of 1~cm$^2$. 
We scan the emulsion plate at 3~r.l. depth 
with the Track Selector, and read out all tracks with ${\tan}{\theta}<0.3$. 
Shower candidates in the track data are identified with off-line programs. 
Selected shower candidates are traced back to the top plate of the chamber 
by searching for tracks having the predicted angle and position. 
We carefully inspect the starting behavior of a shower to 
see whether it was initiated by an electron, gamma ray or hadron. 
For the balloon detectors in 2001, 
we successfully observed atmospheric gamma rays down to 30GeV 
using this automatic scanning system.

\subsubsection{Energy determination}
\begin{figure}[h]
\begin{center}
\includegraphics[width=80mm]{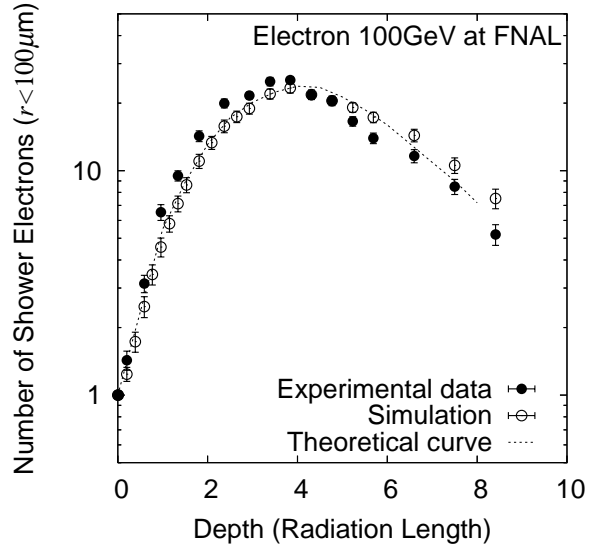}
\end{center}
\caption{\label{fig:fnaltrcv} 
Longitudinal development of the averaged number of shower electrons 
within a radius of 100~$\mu$m. 
Error bars show the uncertainties of the means. 
Theoretical curve is taken from 
Nishimura {\it et al.} (1980) \cite{nishimura80}. }
\end{figure}

\begin{figure}[h]
\begin{center}
\includegraphics[width=80mm]{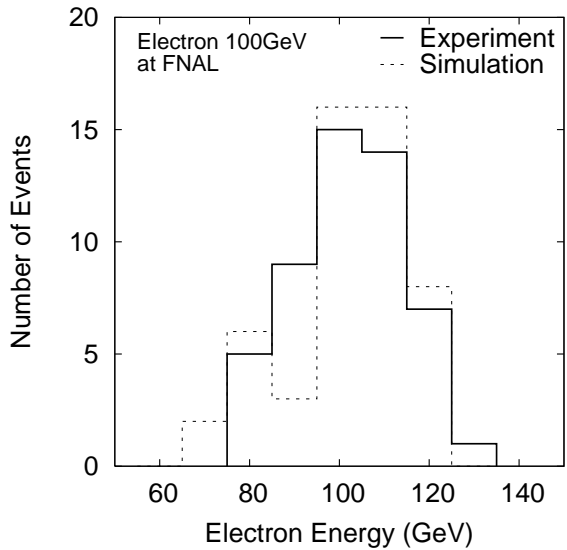}
\end{center}
\caption{\label{fig:fnalhist} 
Energy distributions of the simulated and experimental data 
for 100~GeV electron beams at FNAL. 
}
\end{figure}

\begin{figure}[h]
\begin{center}
\includegraphics[width=40mm]{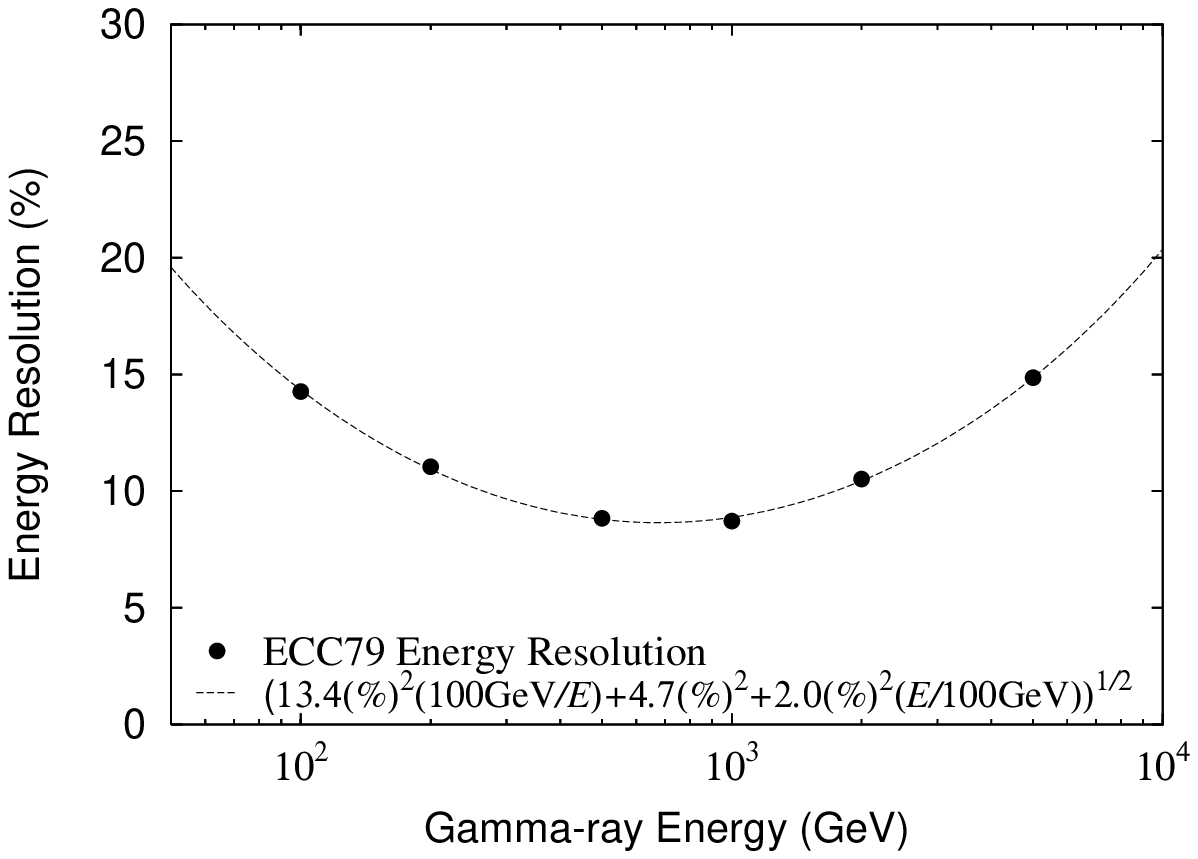}
\includegraphics[width=40mm]{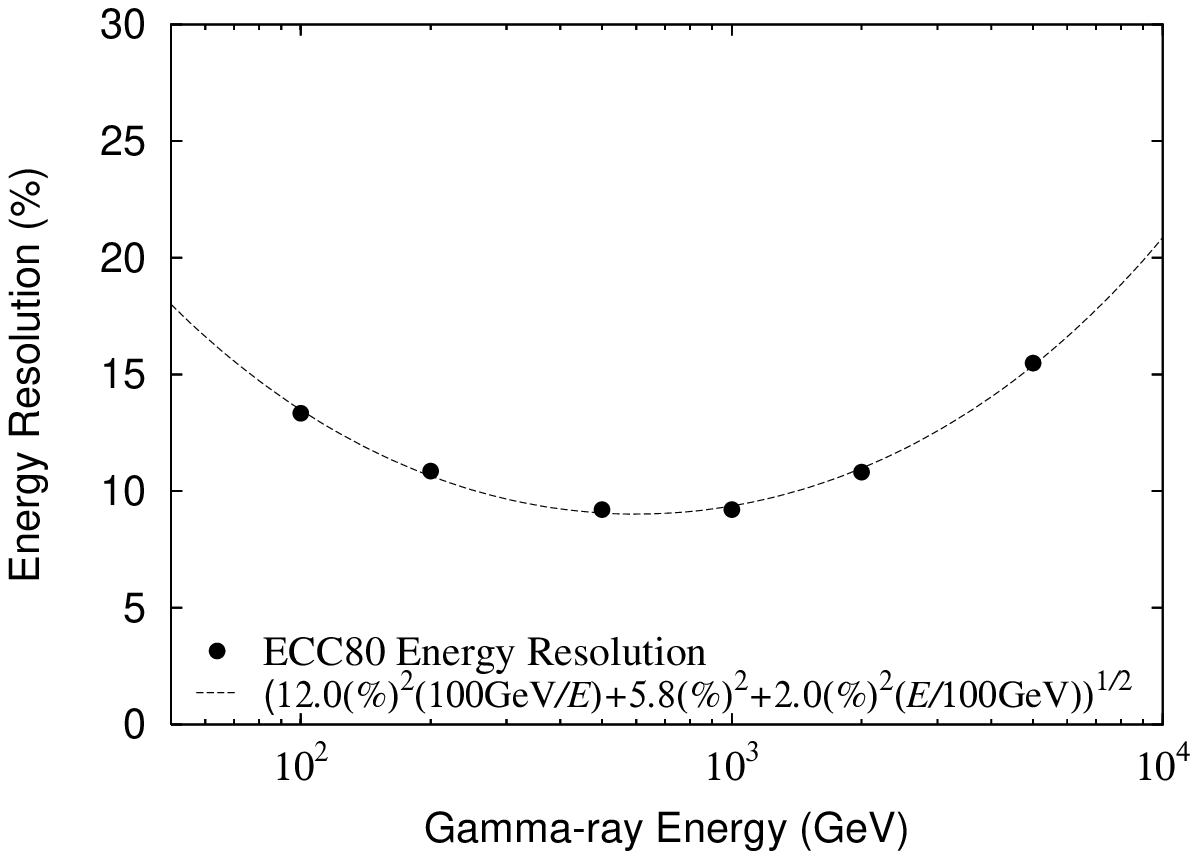}

\includegraphics[width=40mm]{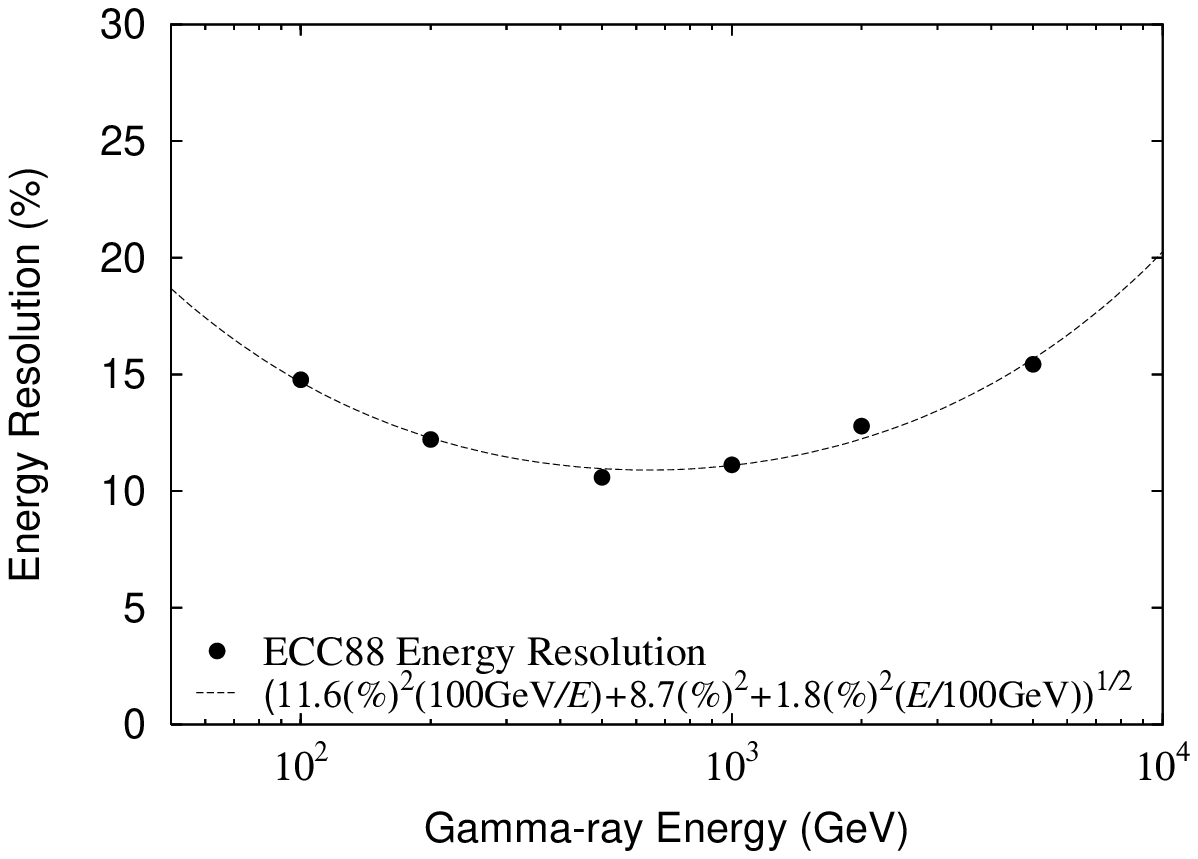}
\includegraphics[width=40mm]{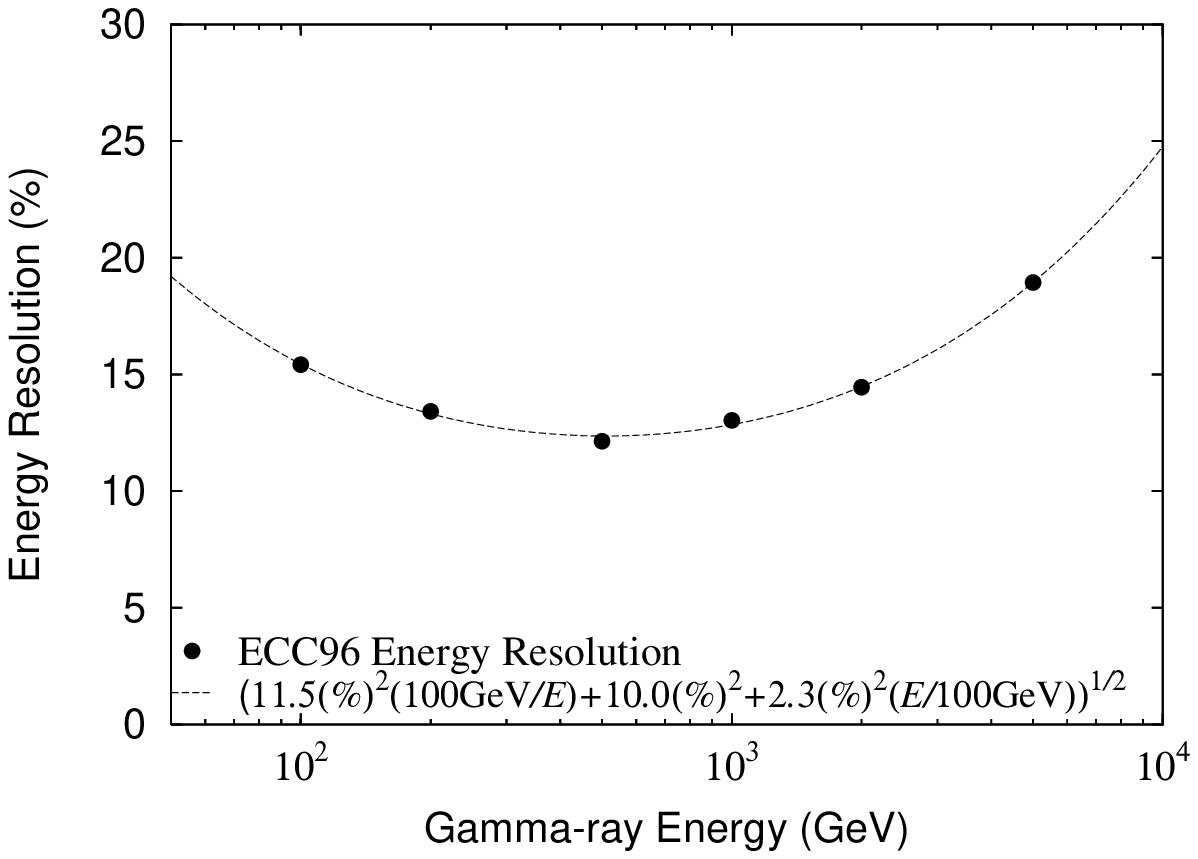}
\end{center}
\caption{\label{fig:eneres}
Examples of the energy dependence of energy resolutions 
with the emulsion chambers for gamma rays from the simulations. 
The dash lines show the fitted function of (\ref{eq:abc}). }
\end{figure}

Shower energy was determined by comparing the number of shower tracks 
at various depths 
with the theoretical transition curves, and fitting to the integrated track length,
used to estimate total ionization in the cascade. 
As the chamber structure is slightly different for each flight, 
we calculated the shower development for each chamber 
using a Monte Carlo simulation code called EPICS \cite{kasahara01}. 
Results calculated using the EPICS code 
were confirmed by emulsion chambers exposed to FNAL 
electron beams \cite{nishimura80} 
in which showers of 100~GeV electrons were re-analyzed. 
Figure~\ref{fig:fnaltrcv},~\ref{fig:fnalhist} 
show longitudinal development of the 
average number of shower electrons, 
and energy distribution of the FNAL experimental and simulated data. 
The simulations well represent the experimental data, 
and the determined energy with the simulation for 100~GeV electrons 
is consistent with the experiment within a precision of a few \%. 
The energy resolution 
is 12~\% at 100~GeV, as shown in Fig.~\ref{fig:fnalhist}. 
The energy resolutions for each emulsion chamber  
are well represented by the form of 
\begin{equation}
 \frac{\sigma}{E_0} = [ a^{2}(\frac{E_0}{100{\rm GeV}})^{-1} + b^{2} + 
	c^{2}(\frac{E_0}{100{\rm GeV}}) ]^{1/2}, 
	\label{eq:abc}
\end{equation}
where $E_0$ is the incident gamma-ray energy and 
$\sigma$ is the standard deviation of energy determination. 
The first term in right-hand side root 
represents statistics-related fluctuations 
of the number of shower particles, while 
the last term represents fluctuations due to ``punch-through", shower particles escaping  
from the finite thickness of the detector. 
The adjustable coefficients of $a$, $b$, and $c$ are derived for each chamber 
using the results of the Monte Carlo simulations. 
Figure \ref{fig:eneres} shows some examples of the energy dependence of 
energy resolution for incident gamma rays from the simulations, 
whose showers start from pair electrons. 
The fitted functions of (\ref{eq:abc}) are also plotted.  
The coefficients of $a$, $b$, and $c$ typically have the values of 
$a{\simeq}12-13~\%$,  $b{\simeq}5-10~\%$,  and $c{\simeq}2-3~\%$.

\subsection{Atmospheric gamma-ray spectrum}

We observed atmospheric gamma rays at each balloon altitude, and 
derived the vertical gamma-ray spectrum normalized at 4.0~g~cm$^{-2}$ 
originated from hadronic interactions using the following formula: 
\begin{multline}
J_{\gamma}(E) = 
( \frac{ N_{\gamma} }{ S{\Omega_{\gamma}}T{\Delta}E C_{\rm eff} C_{\rm enh} }
 - C_{\rm brem} )
{\cdot}C_{\rm alt} \\
({\rm m^{-2} s^{-1} sr^{-1} GeV^{-1}}). 
\label{eq:atmgspec}
\end{multline}
Here $N_{\gamma}$ is the number of gamma-ray events, 
$C_{\rm eff}$ is gamma-ray detection efficiency, 
$C_{\rm enh}$ is enhancement factor due to the energy resolution 
as described below, 
$C_{\rm brem}$ is bremsstrahlung gamma-ray flux from primary electrons, 
and $C_{\rm alt}$ is an altitude conversion factor to 4.0~g~cm$^{-2}$. 
$S{\Omega}_{\gamma}$ is the geometrical factor 
to obtain the vertical flux.

Gamma-ray detection efficiency $C_{\rm eff}$ in the chamber 
is given by 
\begin{equation}
 C_{\rm eff} = 1-\exp(-{\sigma}_0 T_c), 
\end{equation}
where $\sigma_0=0.7733$ is the probability of pair creation in 
one radiation length, 
$T_c$ is the threshold depth of shower starting points 
for gamma rays. 
$C_{\rm eff}$ ranges from 0.902 to 0.955 for our observations, 
corresponding to $T_c = 3.0-4.0$~r.l. in the threshold depth.

The uncertainty of the energy determination has the 
effect of enhancing the absolute flux of gamma rays, in particular,  
for the steep power-law spectrum. 
We derived the enhancement factor $C_{\rm enh}$ due to the energy resolution 
for each chamber, 
which ranges from 1.00 to 1.06 depending on gamma-ray energies and 
emulsion chamber structures 
(see appendix \ref{sec:enhance_factor} in detail).

In this atmospheric gamma-ray spectrum, 
we subtracted the bremsstrahlung gamma rays 
produced by primary cosmic-ray electrons. 
The observed primary electron spectrum $J_{e}(E)$
is well represented by 
\begin{multline}
 J_{e}(E) = 1.6{\times}10^{-4}(E/100{\rm GeV})^{-3.3} \\
  {\rm (m^{-2}s^{-1}sr^{-1}~GeV^{-1})} 
\end{multline}
in the energy range of 30~GeV $-$ 1~TeV \cite{nishimura80}. 
At a depth of $x$~g~cm$^{-2}$, 
the bremsstrahlung gamma-ray spectrum from the electrons 
with a power-law index of $-3.3$ is given by 
\begin{multline}
 C_{\rm brem} = J_{e}(E) C(s=2.3) \\
  \times \frac{e^{-{\sigma_0}{x/X_0}} - e^{-A(s=2.3)x/X_0}}
  {A(s=2.3)-\sigma_0}, 
\end{multline}
where the radiation length in the atmosphere $X_{0}=36.7$~g~cm$^{-2}$, 
$A(s=2.3)=1.674$, and $C(s=2.3)=0.4118$. 
The notations $A$ and $C$ refer to the functions used in 
electro-magnetic shower theory, 
defined in reference \cite{nishimura67}.

The flux of the Galactic diffuse gamma-ray emission 
measured by EGRET 
is less than 0.1~\% of the atmospheric gamma rays 
at depths of several g~cm$^{-2}$ \cite{hunter97}. 
Sreekumar {\it et al.} (1998) \cite{sreekumar98} reported that 
extra-galactic diffuse gamma rays observed with EGRET have a power-law 
spectrum with an index of $-2.1$ in the 50~MeV to 100~GeV. 
Their flux at 100~GeV corresponds to $\sim0.5$~\% 
of the atmospheric gamma rays, and 
there are also arguments that the flux of extra-galactic diffuse gamma rays 
is lower than their derivation \cite{strong04}.  
Therefore, 
we ignored the contribution of the flux of astrophysical gamma rays 
to obtain the atmospheric gamma-ray flux.

The gamma-ray fluxes are normalized to 4.0~g~cm$^{-2}$ equivalent altitude 
from each observation altitude of $x$~g~cm$^{-2}$ using 
\begin{equation}
 C_{\rm alt} = \frac{e^{-4.0/{\Lambda}_p} 
  - e^{-4.0{\sigma_0}/X_0}}{e^{-x/{\Lambda}_p} - e^{-x{\sigma_0}/X_0}}, 
\end{equation}
where ${\Lambda}_p$ is an attenuation length 
of protons in the atmosphere, 
whose value is $\sim100$~g~cm$^{-2}$ in the TeV region 
as described in section \ref{sec:proton}.

\begin{figure}[h]
\begin{center}
\includegraphics[width=80mm]{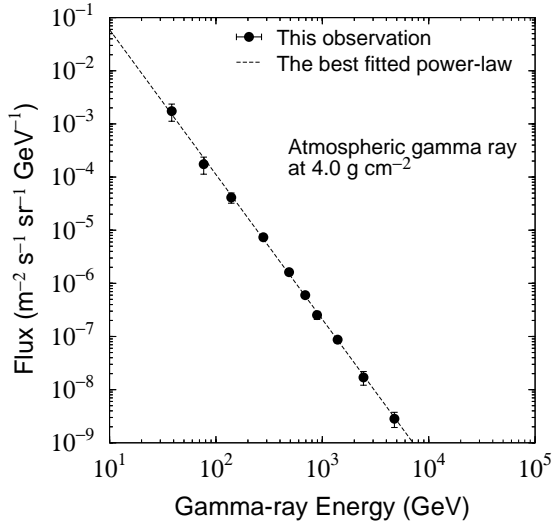}
\end{center}
\caption{\label{fig:gspec} 
The atmospheric gamma-ray spectrum observed 
at an altitude of 4.0~g~cm$^{-2}$. 
The dash line shows the best fit power-law function 
with an index of $-2.73{\pm}0.06$. }
\end{figure}

The total number of observed gamma rays is 330 events in the energy range from  
30~GeV to 8~TeV. 
After the corrections described, 
we derived the vertical spectrum of atmospheric gamma rays. 
Figure \ref{fig:gspec} shows 
the observed spectrum of atmospheric gamma-rays normalized 
at an altitude of 4.0~g~cm$^{-2}$, 
which is well represented by 
\begin{multline}
J_{\gamma}(E)= (1.12\pm0.13)\times10^{-4}(E/100{\rm GeV})^{-2.73\pm0.06} \\
({\rm m}^{-2} {\rm s}^{-1} {\rm sr}^{-1} {\rm GeV}^{-1}). 
\label{eq:fitgspec}
\end{multline}
The flux values and numbers of the gamma rays in each energy bin 
are listed in Table \ref{tab:gamma_flux}.

\begin{table}[h]
\caption{\label{tab:gamma_flux}
Atmospheric gamma-ray fluxes and raw number of gamma rays at 4.0~g~cm$^{-2}$} 
\begin{center}
\begin{tabular}{cccc}
\hline\hline
 Energy    &\ \   $\overline{E}$  &\ \  Raw    &\ \  Flux \\
  (GeV)    &\ \         (GeV)  &\ \  Number &\ \  (m$^{-2}$ s$^{-1}$ sr$^{-1}$ GeV$^{-1}$) \\
\hline
30--50     & $3.84{\times}10^1$  &  8   &  $(1.74{\pm}0.62){\times}10^{-3}$ \\
60--100    & $7.68{\times}10^1$  &  8   &  $(1.75{\pm}0.62){\times}10^{-4}$ \\
100--200   & $1.39{\times}10^2$  & 21   &  $(4.12{\pm}0.90){\times}10^{-5}$ \\
200--400   & $2.79{\times}10^2$  & 59   &  $(7.36{\pm}0.96){\times}10^{-6}$ \\
400--600   & $4.87{\times}10^2$  & 58   &  $(1.62{\pm}0.21){\times}10^{-6}$ \\
600--800   & $6.91{\times}10^2$  & 59   &  $(5.99{\pm}0.78){\times}10^{-7}$ \\
800--1000  & $8.93{\times}10^2$  & 34   &  $(2.53{\pm}0.43){\times}10^{-7}$ \\
1000--2000 & $1.39{\times}10^3$  & 61   &  $(8.70{\pm}1.11){\times}10^{-8}$ \\
2000--3000 & $2.44{\times}10^3$  & 12   &  $(1.70{\pm}0.49){\times}10^{-8}$ \\
3000--8000 & $4.76{\times}10^3$  & 10   &  $(2.84{\pm}0.90){\times}10^{-9}$ \\
\hline
\end{tabular}
\end{center}
\end{table}

\section{\label{sec:proton}Deconvolution of primary proton spectrum}

We assume that the primary protons at the top of atmosphere 
have a power-law spectrum of 
\begin{equation}
 J_{p}(E)dE_{p} = NE_{p}^{-\gamma}dE_{p}.  
\label{eq:pspec}
\end{equation}

As described below, 
the mean free path length of hadronic interactions 
and effective attenuation length of protons in the 
atmosphere are $\lambda_p \simeq 80~{\rm g~cm^{-2}}$ 
and $\Lambda_p \simeq 100~{\rm g~cm^{-2}}$ in the TeV region, 
respectively \cite{mielke94, roesler01}. 
The flux of atmospheric gamma rays observed at 4.0~g~cm$^{-2}$ is 
represented by 
\begin{equation}
 J(E_{\gamma}) = 
 g(E_\gamma) \frac{1}{\lambda_p} 
 \frac{\exp(-4.0/\Lambda_p) - \exp(-4.0\sigma_0/X_0)}
 {\sigma_0/X_0 - 1/\Lambda_p}, 
\label{eq:gspec0}
\end{equation}
considering the absorption of gamma-rays in the atmosphere, 
where $g(E_{\gamma})$ is a production spectrum of gamma-rays 
generated in collisions between protons and atmospheric nuclei.

In the energy range over 100~GeV, 
one can apply a scaling law in the first approximation 
for the production probability of $\pi^0$. 
The $\pi^0$ production rate in a single collision between 
protons and atmospheric nuclei is represented by 
\begin{equation}
 f(E_{\pi^0}/E_{p})d(E_{\pi^0}/E_{p}) 
\label{eq:fspec}
\end{equation}
assuming a scaling law, 
where $E_{\pi^0}$ is the energy of secondary $\pi^0$ 
and $E_{p}$ the energy of parent protons.

\begin{figure}[h]
\begin{center}
\includegraphics[width=80mm]{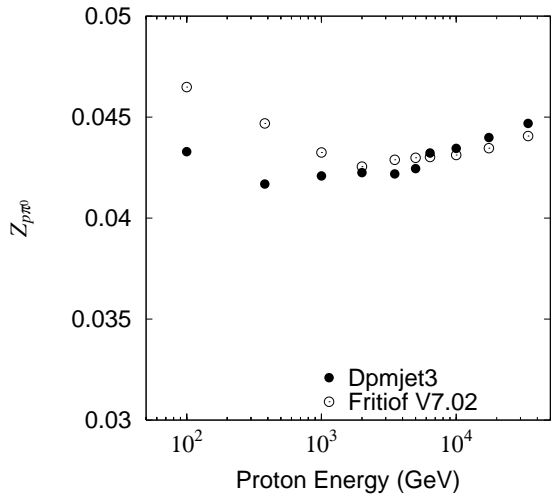}
\end{center}
\caption{\label{fig:z} 
Spectrum weighted moment $Z_{p{\pi^0}}$ of Dpmjet3 and Fritiof V7.02 
for $p+{\rm Air} \rightarrow {\pi^0}+X$ 
with the parent proton energy, 
including the contribution of $\eta$ meson decay 
for gamma rays.}
\end{figure}

The production spectrum of $\pi^0$ generated in collisions 
between protons with a power-law spectrum and atmospheric nuclei 
is now given by 
\begin{equation}
 \Pi(E_{\pi^0})dE_{\pi^0} = 
  \int_{E_{\pi^0}}^{\infty} dE_{p}E_{p}^{-\gamma} 
  f(E_{\pi^0}/E_{p})d(E_{\pi^0}/E_{p}). 
\end{equation}
Putting 
\begin{equation}
 Z_{p{\pi^0}} = \int_{0}^{1}x^{\gamma-1}f(x)dx, 
 \label{eq:zfactor}
\end{equation}
where $x = {E_{\pi^0}}/{E_p}$, 
we have a simple formula of 
\begin{equation}
 \Pi(E_{\pi^0})dE_{\pi^0} = Z_{p{\pi^0}}E_{\pi^0}^{-\gamma}dE_{\pi^0}, 
\label{eq:pispec}
\end{equation}
in which $Z_{p{\pi^0}}$ is the spectrum-weighted moment, 
called as $Z$ factor \cite{gaisserhonda02}. 
We derived a numerical value of $Z_{p{\pi^0}}$ 
for the case of power-law index $\gamma=2.75$, 
using the hadronic interaction models Dpmjet3 \cite{roesler00} 
and Fritiof V7.02 \cite{pi92}. 
The results are shown in Fig.~\ref{fig:z}. 
Here we include the contribution of $\eta$ mesons to gamma rays. 
In implementing hadronic interaction models, 
we used a Monte Carlo simulator called COSMOS\cite{kasahara01}. 
Although we adopted a scaling-law for the deconvolution of 
atmospheric gamma rays to primary protons, 
there is a weak energy dependence in the $Z$ factor of 
the formula (\ref{eq:zfactor}). 
Therefore we need to correct this energy dependence. 
As for the hadronic interaction models used, 
there are several \% differences below 1~TeV 
between Dpmjet3 and Fritiof V7.02, but
above 1~TeV, 
the results from these two models agree well with each other.

\begin{figure}[h]
\begin{center}
\includegraphics[width=80mm]{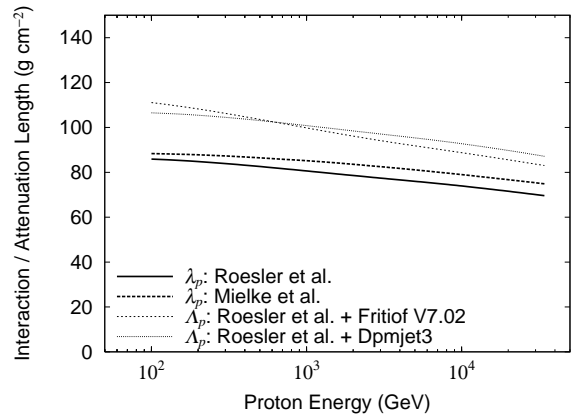}
\end{center}
\caption{\label{fig:int_length} 
Energy dependence of 
mean free path lengths of hadronic interactions $\lambda_p$ 
and attenuation lengths $\Lambda_p$ of protons in the 
atmosphere \cite{mielke94, roesler01}. }
\end{figure}

The production spectrum of gamma rays from a single $\pi^0$ decay 
with energy $E_{\pi^0}$ is given by 
\begin{equation}
 \frac{dn_{\gamma}}{dE_{\gamma}} = \frac{2}{E_{\pi^0}}.
  \label{eq:pi2gspec}
\end{equation}
Combining the formula (\ref{eq:pispec}) and (\ref{eq:pi2gspec}), 
the energy spectrum of gamma rays from a single collision between 
protons and atmospheric nuclei is given by 
\begin{equation}
 g(E) = \int_{E}^{\infty} \frac{dn_{\gamma}}{dE_{\gamma}} \Pi(E_{\pi^0}) 
  dE_{\pi^0}
        = \frac{2Z_{p{\pi^0}}}{\gamma} E^{-\gamma}.
  \label{eq:p2gspec}
\end{equation}
From equation (\ref{eq:gspec0}) and (\ref{eq:p2gspec}), 
the energy spectrum of gamma rays from interactions 
of primary protons in the atmosphere is represented by 
\begin{equation}
 J_{\gamma}(E) = 
  \frac{2}{\gamma} Z_{p \pi^0} C_{\rm atm}^p NE^{-\gamma}, 
  \label{eq:gspec}
\end{equation}
where 
\begin{equation}
 C_{\rm atm}^p = 
  \frac{1}{\lambda_p} 
  \frac{ \exp(-x/\Lambda_p)-\exp(-{\sigma_0} x/X_0) }
  { \sigma_0/X_0 - 1/\Lambda_p }. 
  \label{eq:atm_interact}
\end{equation}
Using this formula, 
we can derive the primary proton spectrum from 
the observed atmospheric gamma rays. 
The interaction length of protons in the atmosphere 
is described by the expression 
$\lambda_p = A {m_p} / {\sigma_p^{air}}\ (A=14.5)$, 
where $A$ is the average mass number of atmospheric nuclei, 
$m_p$ is the mass of protons, and $\sigma_p^{air}$ 
is the cross section for proton-air interaction. 
We also derived the interaction length from the cross section 
given in references \cite{mielke94, roesler01}. 
We derived the attenuation length of protons in the atmosphere 
using the formula 
$\Lambda_p = \lambda_p / (1-Z_{pp}-Z_{pn})$. 
The spectrum-weighted moments of $Z_{pp}$ and $Z_{pn}$ are calculated 
using the hadronic interaction codes of Fritiof V7.02 and Dpmjet3. 
We present the interaction length $\lambda_p$ 
and attenuation length $\Lambda_p$ of protons for various energies 
in Fig.~\ref{fig:int_length}.

Here we need to include 
the contribution due to heavy primaries. 
We corrected for the effects of heavier nuclei such as He, C, N, and O. 
The flux of heavier nuclei  
is ${\sim}6.4$\% of the flux of protons at the 
same energy per nucleon for He, and ${\sim}0.57$\% for C,N,O, 
based on the JACEE and RUNJOB observations \cite{jacee98,runjob01,derbina05}. 
Since interaction lengths for He and C,N,O components are  
${\lambda_{He}}/{\lambda_p} = 0.56$ for He, 
${\lambda_{N}}/{\lambda_p} = 0.29$ for C,N,O, 
using the Hagen-Watts formula \cite{kawamura89}, 
atmospheric interaction contributions are 
$C_{\rm atm}^{He}/C_{\rm atm}^{p} = 1.73$ for He, and 
$C_{\rm atm}^{N}/C_{\rm atm}^{p} = 3.35$ for C,N,O 
from the formula (\ref{eq:atm_interact}). 
According to Dpmjet3, the spectrum-weighted moments are given by  
$Z_{He{\pi^0}}/Z_{p{\pi^0}} = 2.06$ for He and 
$Z_{N{\pi^0}}/Z_{p{\pi^0}} = 4.97$ for C,N,O. 
Therefore, using formula (\ref{eq:gspec}), 
the flux of heavier nuclei can be converted to 
an equivalent proton flux as follows: 
\begin{align}
\frac{J_{He}+J_{CNO}}{J_{p}} &= 
0.064\frac{C_{\rm atm}^{He}}{C_{\rm atm}^{p}}
\frac{Z_{He{\pi^0}}}{Z_{p{\pi^0}}} + 
0.0057\frac{C_{\rm atm}^{N}}{C_{\rm atm}^{p}}
\frac{Z_{N{\pi^0}}}{Z_{p{\pi^0}}} \notag \\
  &= 0.32. 
\end{align}
Therefore, the proton flux from the gamma rays 
is multiplied by $1/1.32=0.76$. 
This correction factor is almost same as the results of an 
independent-nucleon model for heavier nuclei 
as follows: 
$4{\times}0.064+14{\times}0.0057=0.34$, that is $1/1.34=0.75$.  
The uncertainty of the correction factor 
is mainly from the uncertainties of the currently observed flux of 
heavy primary \cite{jacee98,runjob01,derbina05}, 
and is estimated to be no more than ${\sim}5$~\%.

We also corrected for the minor contributions of 
gamma rays from $\eta$ and $K$ mesons, using Dpmjet3 results. 
The contribution
from $\eta$ mesons is 
$\sim0.16$ as large as from $\pi^0$ mesons. 
The correction for $\eta$ mesons 
is included in the calculation of $Z_{p{\pi^0}}$. 
The contribution of gamma rays from $K^0$ mesons is $\sim0.03$ as
large as 
those from $\pi^0$ mesons. 
To correct this contribution, we multiplied the proton flux 
by the correction factor of $1/1.03 = 0.97$.

As shown from equation (\ref{eq:p2gspec}), 
one proton produces gamma-rays of 
$(\gamma/2)(1/Z_{p{\pi^0}})$. 
Hence, the energy of the corresponding parent protons 
is $C_{\rm E} = (2Z_{p{\pi^0}}/\gamma)^{-1/(\gamma-1)}$
times of that of the gamma rays. 
Thus we can presume
that atmospheric gamma rays are on average produced from primary protons 
that have $C_{\rm E}$ times higher energy than the daughter gamma rays. 
$C_{\rm E}$ is ${\sim}7.2$ for Dpmjet3 with $\gamma=2.75$.

\begin{figure}[h]
\begin{center}
\includegraphics[width=80mm]{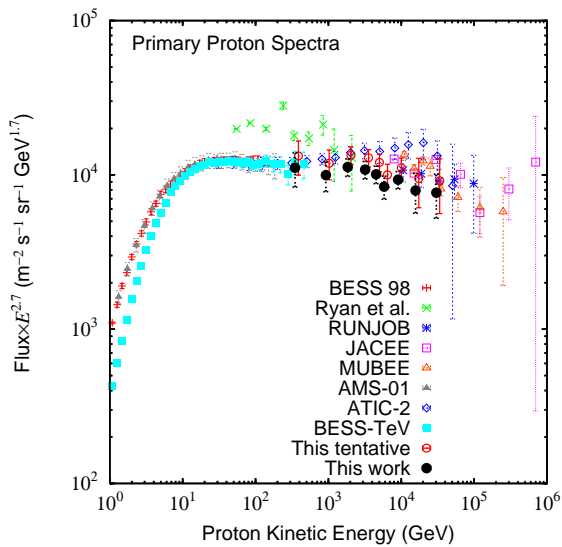}
\end{center}
\caption{\label{fig:pspec} 
The proton spectra deconvolved from the atmospheric gamma-ray 
spectrum, compared with the spectra observed by other groups
\cite{sanuki00,ryan72,runjob01,jacee98,zatsepin93,aguilar02,wefel05,haino04}.
Open circles show the proton spectrum derived from the gamma-ray spectrum with Dpmjet3.  
Solid circles show our final proton spectrum derived with 20~\% larger $Z_{{\pi}^0}$ 
than that of Dpmjet3.}
\end{figure}

In this way, we derived the deconvolved primary proton spectrum as 
\begin{multline}
 J_{p}(E_p) = 
  \frac{\gamma}{2}\frac{1}{Z_{p{\pi^0}}} \lambda_{p} 
  \frac{{\sigma_0}/X_0 - 1/\Lambda_p}
  {\exp(-x/\Lambda_p)-\exp(-{\sigma_0} x/X_0)} \\
  \times 0.76 \times 0.97 \times C_{\rm E}^{-\gamma} J_{\gamma}({E_{\gamma}}),  
  \label{eq:g2pspec}
\end{multline}
where $E_{p} = C_{\rm E}E_{\gamma}$. 
Using this formula, 
we obtained the primary cosmic-ray proton spectrum,
using the Dpmjet3 hadronic interaction model, 
from the atmospheric gamma-ray spectrum at
4.0~g~cm$^{-2}$ as presented in Fig.~\ref{fig:pspec}. 
Figure~\ref{fig:pspec} shows our deconvolved proton spectrum 
with the directly observed proton spectra at high altitude 
obtained by other groups 
\cite{sanuki00,ryan72,runjob01,jacee98,zatsepin93,aguilar02,wefel05,haino04}.  
Our proton flux with Dpmjet3 (open circles) is $\sim20$~\% larger than 
the extrapolation of the BESS, AMS-01, and BESS-TeV data.

The uncertainties in deriving the flux of primary protons 
from atmospheric gamma-ray flux 
come mainly from the hadronic interaction models, 
while the muon flux estimated from the gamma-ray flux 
is less sensitive to the treatment of hadronic interactions, 
as shown in section \ref{sec:muon} and appendix \ref{sec:pion_kaon}. 
For hadronic interactions, 
there are uncertainties in factors such as 
the interaction length of protons in the atmosphere 
and the production rate of pions.  
The former 
has an uncertainty of several \% as shown in Fig.~\ref{fig:int_length}, 
and the uncertainty of the latter is estimated to be much larger, $\sim20-25\%$
\cite{gaisserhonda02}. 
Therefore, 
the main systematic error contribution 
in the estimation of the proton flux from the gamma-ray flux 
comes from the uncertainty of the $Z$ factor for the production rate of pions.  
Sanuki {\it et al.} (2006) pointed out 
that the muon flux measured with the BESS-TeV and L3+C above 100~GeV 
is $\sim20$~\% larger than the calculated muon flux with Dpmjet3, 
and that the muon charge ratio (${\mu}^{+}/{\mu}^{-}$) 
is inconsistent with that predicted by Dpmjet3. 
They suggested that 
in the energy range of 100~GeV to 10~TeV 
the production rates of charged pions and kaons 
are ${\sim}20$~\% larger than that by Dpmjet3 \cite{sanuki06}. 
They also proposed that Dpmjet3 can be modified 
in a phenomenological way, with an assumption 
based on the quark model, to increase by $\sim20$~\% above $\sim100$~GeV 
the Z factors of secondary particles, 
accommodating the calculated muon charge ratio to the observed one.

Referring to their modified model, 
we adopted 20~\% larger $Z_{\pi^0}$ than that of Dpmjet3 and 
derived the proton flux from the gamma-ray flux 
using formula (\ref{eq:g2pspec}). 
The energy of the corresponding parent protons 
is $C_{\rm E}{\simeq}6.5$ times of that of the gamma rays. 
The derived proton spectrum (solid circles) 
is presented in Fig.~\ref{fig:pspec}, and 
is well represented by 
\begin{multline}
J_{p}(E) = (5.4\pm1.2)\times10^{-2}(E/100{\rm GeV})^{-2.79{\pm}0.06} \\
({\rm m}^{-2} {\rm s}^{-1} {\rm sr}^{-1} {\rm GeV}^{-1}) 
\label{eq:fitpspec}
\end{multline}
in the energy range from 200~GeV to 50~TeV. 
The flux values are also summarized in Table~\ref{tab:proton_flux}. 
To derive the proton spectrum, 
we assumed a power-law index of $-2.75$. 
The fitted indexes are $-2.73$ for the gamma rays and 
$-2.79$ for the protons, slightly steeper than the gamma rays,  
because of the weak energy dependences of Z factors and interaction lengths, 
as shown in Fig.~\ref{fig:z} and Fig.~\ref{fig:int_length}. 
The differences of power-law indexes from $-2.75$ 
cause ${\sim}5$~\% differences for the proton flux, 
which are much smaller than the main uncertainty of hadronic interaction models 
of ${\sim}20$~\%.

Our deconvolved proton spectrum above 10~TeV is somewhat smaller than 
the JACEE and MUBEE data, 
and agrees well with the RUNJOB data. 
The ATIC-2 data show much higher flux than our proton flux. 
Below 10~TeV, our proton spectrum is consistent with 
the extrapolated fluxes with the BESS, AMS-01, and BESS-TeV data.

\begin{table}[h]
\caption{\label{tab:proton_flux}
Deconvolved primary proton fluxes} 
\begin{center}
\begin{tabular}{cc}
\hline\hline
$\overline{E}$  &\ \  Flux \\
(GeV)           &\ \  (m$^{-2}$ s$^{-1}$ sr$^{-1}$ GeV$^{-1}$) \\
\hline
$3.49{\times}10^2$  &  $(1.52{\pm}0.38){\times}10^{-3}$ \\
$9.23{\times}10^2$  &  $(9.83{\pm}2.15){\times}10^{-5}$ \\
$1.84{\times}10^3$  &  $(1.72{\pm}0.22){\times}10^{-5}$ \\
$3.20{\times}10^3$  &  $(3.73{\pm}0.48){\times}10^{-6}$ \\
$4.52{\times}10^3$  &  $(1.36{\pm}0.18){\times}10^{-6}$ \\
$5.83{\times}10^3$  &  $(5.72{\pm}0.97){\times}10^{-7}$ \\
$9.04{\times}10^3$  &  $(1.94{\pm}0.25){\times}10^{-7}$ \\
$1.57{\times}10^4$  &  $(3.72{\pm}1.07){\times}10^{-8}$ \\
$3.03{\times}10^4$  &  $(6.07{\pm}1.92){\times}10^{-9}$ \\
\hline
\end{tabular}
\end{center}
\end{table}

\section{\label{sec:muon}Estimation of high-energy muon spectrum 
at high altitude}

Atmospheric muons mainly come from the decay of charged pions and kaons. 
We discuss the contributions of pions and kaons for muon production.

In the first approximation, the same number of 
$\pi^{-}$ and $\pi^{+}$ mesons are produced as
$\pi^0$. 
Therefore, 
the production spectrum of charged pions can be estimated from 
that of atmospheric gamma rays through $\pi^0$, 
independent of primary cosmic rays, such as 
$F_{{\pi}^{\pm}} \simeq 2F_{\pi^0} \simeq {\gamma}F_{\gamma}$, 
where $F_{\pi^{\pm}}$, $F_{\pi^0}$ and $F_{\gamma}$ 
are the production spectrum of $\pi^{\pm}$, $\pi^0$ and 
atmospheric gamma rays per unit depth, respectively. 
In more detail, 
the production ratio of charged pions to neutral pions 
is calculated to be $1.7$ using a hadronic interaction model of Dpmjet3.

Since the $\pi^{\pm}$ lifetime for decay to muons is 
$2.6\times10^{-8}$~sec, 
$\pi^{\pm}$ mesons above $\sim$100~GeV 
travel more than $\sim$5~km on the average and may
interact with atmospheric nuclei before decay. 
On the other hand, 
charged kaons have larger mass, 493.6~MeV, and shorter life time, 
$1.2\times10^{-8}$~sec. 
Therefore, 
although atmospheric muons are mainly produced by the decay of $\pi^{\pm}$, 
the relative contribution of kaons to the muon flux increases
with increasing energy above $\sim$100~GeV \cite{gaisserhonda02}.

\begin{figure}[h]
\begin{center}
\includegraphics[width=80mm]{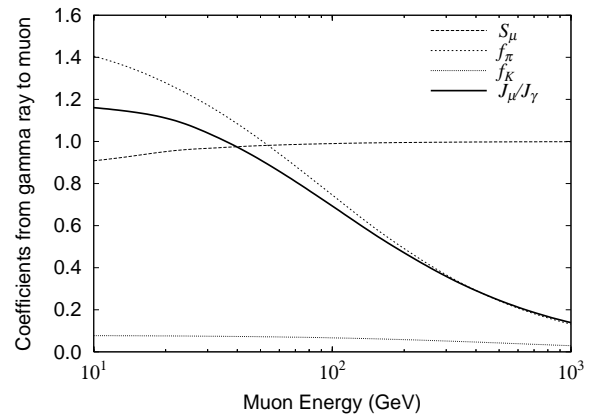}
\end{center}
\caption{\label{fig:g2mu_coeff}
Coefficients describing the relation between 
the muon flux $J_{\mu}$ and the gamma-ray flux $J_{\gamma}$. 
$J_{\mu} = S_{\mu} (f_{\pi^{\pm}}+f_K)J_{\gamma}/(1+f_{\eta})$, 
where $f_{\eta} = 0.16$. 
$S_{\mu}$ is a survival probability of ${\mu}^{\pm}$. 
$f_{\pi^{\pm}}$, $f_K$, and $f_{\eta}$ are flux ratios of 
${\pi}^{\pm}{\rightarrow}{\mu}^{\pm}$ to ${\pi}^{0}{\rightarrow}{\gamma}$, 
${K}^{\pm}{\rightarrow}{\mu}^{\pm}$ to ${\pi}^{0}{\rightarrow}{\gamma}$, 
and 
${\eta},K^{0}{\rightarrow}{\gamma}$ to ${\pi}^{0}{\rightarrow}{\gamma}$, 
respectively. 
See text in detail.}
\end{figure}

Hence, 
on the basis of the gamma-ray flux from $\pi^0$, 
we can derive the flux of muons from $\pi^{\pm}$ and $K$, 
correcting for muon decay and 
the flux of gamma rays from $\eta$ and $K^0$ mesons. 
The muon flux is represented as the following formula of 
\begin{equation}
 J_{\mu} = S_{\mu} (f_{\pi^{\pm}}+f_{K}) \frac{J_{\gamma}}{1+f_{\eta}},
 \label{eq:g2mu}
\end{equation}
where 
$S_{\mu}$ is the survival probability of muons, 
$f_{\pi^{\pm}}$ is the ratio of the muon flux from $\pi^{\pm}$ 
to the gamma-ray flux from $\pi^0$, 
$f_{K}$ is a ratio of the muon flux from kaons to 
the gamma-ray flux from $\pi^0$, and 
$f_{\eta}$ is the ratio of the gamma-ray flux from $\eta+K$ to 
the gamma-ray flux from $\pi^0$. 
We derived the production ratio of $K^{\pm}/{\pi}^{\pm}$ using 
the hadronic interaction model of Dpmjet3. 
Figure \ref{fig:g2mu_coeff} shows 
$S_{\mu}$, $f_{\pi}$, $f_{K}$, and $f_{\eta}$ as a function of muon energy. 
The detailed derivation for these parameters is given 
in appendix \ref{sec:pion_kaon}.

\begin{figure}[h]
\begin{center}
\includegraphics[width=80mm]{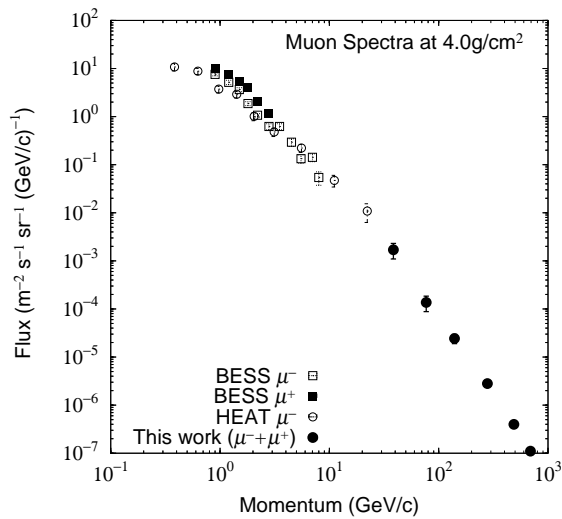}
\end{center}
\caption{\label{fig:muspec} 
 Comparison of muon spectra estimated by our gamma-ray observation with 
 those observed by BESS\cite{sanuki01} and HEAT\cite{beatty04}. 
 The altitudes are normalized at 4.0~g~cm$^{-2}$. 
 In the direct observations charged muons are measured individually, 
 and in our estimation all charged muons are included together.}
\end{figure}

We transformed the observed atmospheric gamma-ray spectrum to the muon spectrum 
using the formula (\ref{eq:g2mu}), 
and results are presented in Fig.~\ref{fig:muspec} and Table~\ref{tab:muon_flux}.  
In Fig.~\ref{fig:muspec}, 
we compare our estimated $\mu^{+} + \mu^{-}$ spectrum at 4.0~g~cm$^{-2}$ 
with that of $\mu^+$ and $\mu^-$ 
observed by BESS and HEAT group \cite{sanuki01,beatty04} as representative values, 
because the results by the BESS, HEAT, MASS and CAPRICE 
are consistent with each other within statistical errors. 
Since the BESS and HEAT observations are performed at altitudes of 
4.5~g~cm$^{-2}$ and 5.6~g~cm$^{-2}$ respectively, 
the coefficients of $4.0/4.5$ and $4.0/5.6$ are multiplied to obtain 
the muon flux at 4.0~g~cm$^{-2}$ for the comparison. 
As shown in Fig.~\ref{fig:muspec}, our
flux is just on the line of extrapolation of 
each datum of the BESS and HEAT experiments and 
gives consistent values within statistical errors, 
although the energy region of our gamma rays is 
almost one order of magnitude higher than 
that of the directly observed muons.

To calculate the atmospheric neutrino flux at energies above $\sim100$~GeV 
with better accuracy, 
muon flux data above $\sim100$~GeV measured at high altitude 
would be very useful for the calibration of the hadronic interaction model. 
However, 
we have no direct observations of muons above $\sim100$~GeV at high altitude. 
Our estimated muon flux gives useful information for this purpose.

\begin{table}[h]
\caption{\label{tab:muon_flux}
Estimated atmospheric ${\mu}^{+} + {\mu}^{-}$ fluxes at 4.0~g~cm$^{-2}$} 
\begin{center}
\begin{tabular}{cc}
\hline\hline
$\overline{P}$  &\ \  Flux \\
(GeV/c)           &\ \  (m$^{-2}$ s$^{-1}$ sr$^{-1}$ (GeV/c)$^{-1}$) \\
\hline
$3.85{\times}10^1$  &  $(1.70{\pm}0.61){\times}10^{-3}$ \\
$7.69{\times}10^1$  &  $(1.36{\pm}0.48){\times}10^{-4}$ \\
$1.39{\times}10^2$  &  $(2.43{\pm}0.53){\times}10^{-5}$ \\
$2.79{\times}10^2$  &  $(2.80{\pm}0.37){\times}10^{-6}$ \\
$4.87{\times}10^2$  &  $(3.97{\pm}0.52){\times}10^{-7}$ \\
$6.91{\times}10^2$  &  $(1.10{\pm}0.14){\times}10^{-7}$ \\
\hline
\end{tabular}
\end{center}
\end{table}

\section{\label{sec:summary}Summary}

We have observed atmospheric gamma rays at several g~cm$^{-2}$ altitude 
with emulsion chambers. 
Precise measurements of the atmospheric gamma-ray spectrum 
from hadronic interactions have been performed 
with the largest $S{\Omega}T$ among all existing measurements 
in the energy range 30~GeV to 8~TeV. 
To obtain the atmospheric gamma-ray flux at 4.0~g~cm$^{-2}$ 
from the observed data at each altitude, 
we took into account several correction factors such as 
gamma-ray detection efficiency, enhancement due to energy resolution, 
bremsstrahlung gamma-ray flux from primary electrons, 
and altitude conversion to 4.0~g~cm$^{-2}$. 
Although some electronic detectors have uncertainty in the determination of 
their geometrical factor $S{\Omega}_{\gamma}$, 
that of the emulsion chambers 
can be estimated very accurately 
because of the precise determination of the track location and 
the simple configuration of the detector. 
Thus, the uncertainties of these correction factors and $S{\Omega}_{\gamma}$ 
are relatively smaller than the statistical errors. 
We deconvolved primary proton spectrum 
in the 200~GeV $-$ 50~TeV from our gamma-ray spectrum in a reliable way, 
assuming 
a single interaction of each proton with an atmospheric nucleus. 
The main uncertainty in the deconvolution 
comes from hadronic interaction models, and 
we referred to a phenomenologically modified model in Dpmjet3 \cite{sanuki06}. 
While in the energy range from 100~GeV to 10~TeV 
accurate data are missing in the currently observed proton spectra, 
our estimated proton spectrum fills this gap. 
Our derived proton flux is consistent with the other observed data 
in the overlapping region. 
This may also indicate the validity of the hadronic interaction model 
in the TeV region proposed by Sanuki {\it et al.} \cite{sanuki06}. 
From the gamma-ray spectrum, 
we also deconvolved the atmospheric muon spectrum,  
which is consistent with direct muon observations below 10~GeV.

\begin{acknowledgments}
We sincerely thank the crews of the SBC of JAXA/ISAS and the NSBF 
for the successful balloon flights.
We thank K.Kasahara and M.Honda for their helpful discussions 
and help in implementing hadronic interaction codes. 
We are also grateful to R.J.Wilkes for his careful reading of the manuscript. 
\end{acknowledgments}

\appendix

\section{Enhancement factor of flux by energy resolution}
\label{sec:enhance_factor}

Defining $E_0$ as the incident gamma-ray energy and 
$\sigma$ as the standard deviation of energy determination, 
the observed energy $E$ has an uncertainty given by 
the Gaussian probability function
\begin{equation}
 P(E-E_0) = \frac{1}{\surd\overline{2\pi}\sigma} 
  \exp(-(E-E_0)^2 / (2\sigma^2)). 
  \label{eq:gaus_prob}
\end{equation}
As the gamma-ray spectrum is a power-law function of $E_0^{-\gamma} dE_0$,  
the enhancement factor is given by 
\begin{equation}
 C_{\rm enh} = \frac{1}{E^{-\gamma}} \int_0^{\infty} E_0^{-\gamma}P(E-E_0) dE_0. 
 \label{eq:obs_powlaw}
\end{equation}
In the case of constant energy resolution $\sigma/E_0={\rm const.}$, 
it is well represented by the series form of 
\begin{equation}
C_{\rm enh} = 1 + \frac{(\gamma-1)(\gamma-2)}{2}(\frac{\sigma}{E_0})^2 + \cdots, 
\end{equation}
independent of the gamma-ray energy. 
The exact solution can be presented by a hypergeometric function.  
In the case of energy resolution $\sigma/E_0=15\%$, with $\gamma=2.75$, 
this enhancement factor is 1.01. 
In emulsion chambers, 
the energy resolution is well represented by the form of (\ref{eq:abc}). 
Using numerical integration with the formula (\ref{eq:gaus_prob}), 
(\ref{eq:obs_powlaw}), and (\ref{eq:abc}), 
we derived the enhancement factor $C_{\rm enh}$ for each gamma-ray energy bin 
and for each emulsion chamber. 
$C_{\rm enh}$ has values from 1.00 to 1.06.

\section{Contribution of pions and kaons for muon production}
\label{sec:pion_kaon}

\subsection{Decay factor $B$ for each particle}

In the case of an isothermal atmosphere, 
secondary particles produced at a depth of $x$~g~cm$^{-2}$
arrive at a depth of $x_0$~g~cm$^{-2}$ without decay  
with a survival probability of $(x/x_0)^{(B/E)}$ per unit depth, 
neglecting the ionization loss of the particles, 
where $B$ is the decay factor and $E$ is the energy for each particle 
\cite{hayakawa69}. 
We define the decay factor $B$ as $B = Hm/({\tau}c)$, 
where $H$ is the scale height of the atmosphere, 
$m$ is particle mass, 
$\tau$ is particle decay lifetime, 
and $c$ is light speed. 
Setting scale height $H$ to be 6.3~km  at balloon altitude,  
the decay factors are $B_{\mu}=1.0$~GeV for muons, 
$B_{\pi}=112.7$~GeV for pions, and  $B_{K}=838.5$~GeV for kaons.

\subsection{Decay of muons}

At balloon altitudes of several g~cm$^{-2}$ residual overburden, 
atmospheric muons are produced at a rate proportional to 
the transverse depth $x$, 
because the interaction length for cosmic rays on 
atmospheric nuclei is much larger than the atmospheric depth of several g~cm$^{-2}$. 
Therefore, 
the survival probability $S_{\mu}$ of muons without decay is given 
by 
\begin{equation}
 S_{\mu} = \frac{1}{x_0} \int_{0}^{x_0} (\frac{x}{x_0})^{B_{\mu}/E} dx  
 = \frac{E}{B_{\mu}+E}. 
 \label{eq:smu}
\end{equation}

\subsection{Decay of pions}

The flux of $\pi^{\pm}$ decaying per unit depth is obtained by 
multiplying $B_{\pi}/E$ into the formula (\ref{eq:smu}) 
and normalizing by the production rate as 
\begin{equation}
 (\frac{B_{\pi}}{B_{\pi}+E})F_{\pi}. 
\end{equation}
Hence the flux of $\pi^{\pm}$ decaying until a depth of $x_0$~g~cm$^{-2}$ 
is given by 
\begin{equation}
 x_0(\frac{B_{\pi}}{B_{\pi}+E})F_{\pi}
 = 0.85{\gamma}(\frac{B_{\pi}}{B_{\pi}+E})J_{\gamma}, 
\end{equation}
where $J_{\gamma}$ is the atmospheric gamma-ray flux at $x_0$~g~cm$^{-2}$ 
and given by $J_{\gamma} = x_{0} F_{\gamma}$ 
from the relation of $F_{\pi}=1.7F_{\pi^0}=0.85{\gamma}F_{\gamma}$. 
Here, 
the production ratio of $\pi^{\pm}$ to $\pi^0$ is calculated to be 1.7 
using the hadronic interaction code Dpmjet3. 

As described in reference \cite{hayakawa69}, 
the flux of muons produced from the decay of pions 
at a depth of $x$~g~cm$^{-2}$, per unit depth, is given by 
\begin{equation}
 \frac{m_{\pi}^2}{m_{\pi}^2-m_{\mu}^2} \int_{E_{-}}^{E_{+}} 
 (\frac{B_{\pi}}{B_{\pi}+E_{\pi}})F_{\pi}(E_{\pi}) \frac{dE_{\pi}}{E_{\pi}}, 
\end{equation}
where $E_{+}$ and $E_{-}$ show the upper and lower limits of the energy 
of pions with rest mass $m_{\pi}$  
that produce muons with rest mass $m_{\mu}$ 
energy $E_{\mu}$. 
$E_+$ and $E_-$ are given by 
\begin{equation}
 E_{+} = (\frac{m_{\pi}}{m_{\mu}})
 (\frac{E_{\mu}E_{\mu}^{*}+p_{\mu}p_{\mu}^{*}c^{2}}{m_{\mu}c^2})
\simeq 
(\frac{m_{\pi}}{m_{\mu}})^2 E_{\mu} = \frac{E_{\mu}}{r_{\pi}}
\end{equation}
and 
\begin{equation}
 E_{-} = (\frac{m_{\pi}}{m_{\mu}})
 (\frac{E_{\mu}E_{\mu}^{*}-p_{\mu}p_{\mu}^{*}c^{2}}{m_{\mu}c^2})
\simeq E_{\mu},  
\end{equation}
where $r_{\pi}$ is $(m_{\mu}/m_{\pi})^2 = 0.5733$, 
$p_{\mu}$ is momentum of pion, and 
the notation '$*$' means a static system.

In the case of the complete decay of pions, 
the flux of muons per unit depth 
at a depth of $x_0$~g~cm$^{-2}$ is given by 
\begin{multline}
x_0 \frac{m_{\pi}^2}{m_{\pi}^2-m_{\mu}^2} 
\int_{E_{-}}^{E_{+}}  F_{\pi}(E_{\pi}) \frac{dE_{\pi}}{E_{\pi}} \\
= \frac{x_0}{1-r_{\pi}} \int_{E_{\mu}}^{E_{\mu}/r_{\pi}} 
  E_{\pi}^{-\gamma} \frac{dE_{\pi}}{E_{\pi}} 
= x_0 \frac{1-r_{\pi}^{\gamma}}{1-r_{\pi}}
 \frac{E_{\mu}^{-\gamma}}{0.85\gamma}. 
\end{multline}

As a result, 
the ratio of the flux of muons from the decay of charged pions   
to the flux of gamma rays from the decay of neutral pions is given by 
\begin{equation}
 f_{\pi}(E_{\mu}) = \frac{ 0.85{\gamma}E_{\mu}^{\gamma} }{ 1-r_{\pi}^{\gamma} } 
  \int_{E_{\mu}}^{E_{\mu}/r_{\pi}}  ( \frac{ B_{\pi} }{ B_{\pi}+E_{\pi} } )
  E_{\pi}^{-\gamma} \frac{dE_{\pi}}{E_{\pi}}. 
\end{equation}

\subsection{Decay of kaons}

Although muons are mainly produced from the decay of charged
pions, 
there is a minor contribution from kaons \cite{gaisserhonda02}. 
While there are $K^{\pm}$, $K^0_S$ and $K^0_L$,  
only the charged kaons decay directly to muons via 
$K^{\pm} \rightarrow \mu^{\pm}+\nu$. 
For the production of muons 
the contribution of kaons can be treated in the same way as pions. 
We derive the relation between kaons and gamma rays 
through the production ratio of $K^{\pm}/{\pi^{\pm}}$. 
The production ratio of $K^{\pm}/{\pi^{\pm}}$ is 0.14 
in the parent proton energy of 200~GeV $-$ 40~TeV 
according to the hadronic interaction code in Dpmjet3. 
The decay mode $K^{\pm}{\rightarrow}{\mu}^{\pm}$ 
dominates (63~\% of decays) among the charged kaon decay mode \cite{pdg04}. 
As a result, 
we can obtain the ratio of the flux of muons from kaons 
to the flux of gamma rays from neutral pions as 
\begin{multline}
 f_{K}(E_{\mu}) = 0.14{\times}0.63{\times}
 \frac{ 0.85{\gamma}E_{\mu}^{\gamma} }{ 1-r_{K}^{\gamma} } \\
 \int_{E_{\mu}}^{E_{\mu}/r_{K}} (\frac{ B_{K} }{ B_{K}+E_{K} })
 E_{K}^{-\gamma} \frac{dE_{K}}{E_{K}}, 
\end{multline}
where $r_{K}$ is $(m_{\mu}/m_{K})^2 = 0.0459$. 
Although there is a $K^{\pm} \rightarrow \pi^{\pm}$ decay mode (21~\% of decays), 
the contribution of 
$K^{\pm} \rightarrow \pi^{\pm} \rightarrow \mu^{\pm}$ 
is less than $\sim$1~\% for the total muon flux.

Since kaons have larger rest mass and shorter life time, 
the decay factor of kaons, $B_K = 838.5$~GeV, 
becomes larger than that of pions, $B_{\pi} = 112.7$~GeV.  
Therefore, 
the contribution of kaons continues into a higher energy region 
than pions.

\bibliography{atm_gamma}

\end{document}